\documentclass[pra,twocolumn]{revtex4}
\usepackage{amssymb}
\usepackage{amsfonts}
\usepackage{amsmath}
\usepackage{graphicx}

\begin{document}

\title{Quantum non-Markovian \textquotedblleft casual
bystander\textquotedblright\ environments}
\author{Adri\'{a}n A. Budini}
\affiliation{Consejo Nacional de Investigaciones Cient\'{\i}ficas y T\'{e}cnicas
(CONICET), Centro At\'{o}mico Bariloche, Avenida E. Bustillo Km 9.5, (8400)
Bariloche, Argentina, and Universidad Tecnol\'{o}gica Nacional (UTN-FRBA),
Fanny Newbery 111, (8400) Bariloche, Argentina}
\date{\today }

\begin{abstract}
Quantum memory effects can be induced even when the degrees of freedom
associated to the environment are not affected at all during the system
evolution. In this paper, based on a bipartite representation of the
system-environment dynamics, we found the more general interactions that
lead to this class of quantum non-Markovian \textquotedblleft casual
bystander environments.\textquotedblright\ General properties of the
resulting dynamics are studied with focus on the system-environment
correlations, a collisional measurement-based representation, and the
quantum regression hypothesis. Memory effects are also characterized through
an operational approach, which in turn allows to detect when the studied
properties apply. Single and multipartite qubits dynamics support and
exemplify the developed results.
\end{abstract}

\maketitle

\section{Introduction}

Both in classical and quantum realms, memory effects emerge whenever a set
of dynamical degrees of freedom is not considered as part of the system of
interest \cite{vanKampen,breuerbook,vega,wiseman}. In classical systems (or
incoherent ones), the presence of memory effects can be related to
departures from a Markov property defined in a probabilistic frame. In
contrast, the definition of memory effects and non-Markovianity is much more
subtle in a quantum regime.

Given that any quantum system is affected by a measurement process, a
reasonable approach to defining quantum non-Markovianity is to study the
properties of the (unperturbed) system density matrix propagator. In fact,
the theory of quantum semigroups \cite{alicki}\ is usually taken as a
landmark of quantum Markovianity. Thus, any departure of the propagator
properties with respect to that of a Lindblad evolution can be proposed as a
signature of quantum non-Markovianity \cite{BreuerReview,plenioReview}. This
approach has shown to be a very fruitful tool to study memory effects in
quantum systems, leading to the formulation of many, in general
inequivalent, memory witnesses (see for example Refs.~\cite%
{BreuerFirst,EnergyBackFLow,Energy,HeatBackFLow,rivas,DarioSabrina,canonicalCresser}%
). One interesting perspective that these studies provide is the
understanding of memory effects through an \textquotedblleft
environment-to-system backflow of information\textquotedblright ~\cite%
{BreuerFirst,EnergyBackFLow,Energy,HeatBackFLow}. Information stored in the
environment degrees of freedom may influence the system at later times,
giving a solid and clear understanding of memory effects.

In spite of the simplicity and efficacy\ of the previous theoretical
perspective, it has been shown that memory effects may emerge even when the
degrees of freedom associated to the environment are not affected at all
during the system evolution, which implies the absence of any
\textquotedblleft physical\textquotedblright\ environment-to-system backflow
of information. A clear situation where this occurs is in\ quantum systems
coupled to incoherent degrees of freedom that have a fixed classical
stochastic dynamics \cite{megier,maximal}.

While the previous drawback in the definition of information flows remains
under debate \cite%
{megier,maximal,petruccione,amato,santis,acin,horo,EntroBack}, alternative
operational approaches to quantum memory effects \cite%
{modi,pollock,pollockInfluence,goan,budiniCPF,budiniChina,bonifacio,han,BIF,ban}
furnish a possible solution. For example, by subjecting a system to three
successive measurements, a conditional past-future correlation \cite%
{budiniCPF}\ provides a memory witness that is consistent with the usual
probabilistic approach to non-Markovianity. This object is defined by the
correlation between the first and last outcomes conditioned to a given
intermediate value. It vanishes in a (probabilistic) Markovian regime. In
addition, by randomizing the intermediate post-measurement system state,
even in presence of memory effects, the conditional past-future correlation
vanishes when the environment is not affected by the system evolution \cite%
{BIF}. Thus, it detects the presence or absence of \textquotedblleft
bidirectional system-environment information flows.\textquotedblright\ This
last property provides a solution to the previous drawback. In fact, a solid
experimental procedure for distinguishing between memory effects induced by
environments that are affected or not by their interaction with the system
is established.

The previous advancements left open an interesting issue whose formulation
is \textit{independent} of any memory witness definition. Besides
environments consisting of incoherent degrees of freedom with a fixed
classical stochastic dynamics, in which other situations it may occur that
the environment is not affected at all by its interaction with the system?
More specifically, we ask about the most general system-environment
interactions that guarantee this property. The resolution of this problem is
relevant for achieving a clear understanding of intrinsically different
memory effects, that is, those where the environment is or not altered by
its interaction with the system.

The main goal of this work is to answer the previous question. We
characterize the most general system-environment interactions that, even
when the system develops memory effects, guarantee that the environment
self-dynamics and state remain independent of the system degrees of freedom.
For these \textquotedblleft casual bystander
environments,\textquotedblright\ special interest is paid to the resulting
system dynamics, the system-environment correlations, a collisional
measurement-based representation \cite%
{embedding,collisionVacchini,ciccarello,palmaMultipartito,strunz,portugal,brasilCollisional,brasil}%
, and the validity of the quantum regression hypothesis when analyzing
system operator correlations \cite{QRTVacchini,QRTLindbladRate,mutual,banQRT}%
. General expressions for the conditional past-future correlation\ \cite%
{budiniCPF,BIF} are also provided. The main results are exemplified by
studying these properties and indicators for a class of qubit dynamics with
single and multipartite dephasing channels.

The paper is outlined as follows. In Sec. II we derive the most general
interaction consistent with a causal bystander environment. General
properties of the resulting system dynamics are analyzed in Sec. III. In
Sec. IV we study single and multipartite examples. The conclusions are
provided in Sec. V. Calculus details are presented in the Appendix.

\section{Quantum \textquotedblleft casual bystander\textquotedblright\
environments}

We consider a system ($s$) interacting with uncontrollable quantum degrees
of freedom that constitute the environment ($e$). Correspondingly, in the
total Hilbert space $\mathcal{H}_{s}\otimes \mathcal{H}_{e},$ the bipartite
density matrix is denoted as $\rho _{t}^{se}.$ Its evolution is written as%
\begin{equation}
\frac{d}{dt}\rho _{t}^{se}=(\mathcal{L}_{s}+\mathcal{L}_{e}+\mathcal{L}%
_{se})[\rho _{t}^{se}],  \label{RhoSUEvolution}
\end{equation}%
where $\mathcal{L}_{s}$ and $\mathcal{L}_{e}$ define the system and
environment isolated dynamics respectively, while $\mathcal{L}_{se}$
introduce their mutual interaction. As usual, the marginal system and
environment states follow by tracing out the complementary degrees of
freedom,%
\begin{equation}
\rho _{t}^{s}=\mathrm{Tr}_{e}[\rho _{t}^{se}],\ \ \ \ \ \ \ \ \ \ \ \rho
_{t}^{e}=\mathrm{Tr}_{s}[\rho _{t}^{se}],  \label{Rhoes}
\end{equation}%
where $\mathrm{Tr}[\cdots ]$ is the trace operation. By \textit{definition},
a casual bystander environment is characterized by a density matrix $\rho
_{t}^{e}$ that is completely independent of the system state and dynamics.
Consequently, its time-evolution $(d\rho _{t}^{e}/dt)$ must also fulfill the
same property. From Eq.~(\ref{RhoSUEvolution}) we get $(d\rho _{t}^{e}/dt)=%
\mathcal{L}_{e}[\rho _{t}^{e}]+\mathrm{Tr}_{s}(\mathcal{L}_{se}[\rho
_{t}^{se}]),$ leading to the condition%
\begin{equation}
\mathrm{Tr}_{s}(\mathcal{L}_{se}[\rho _{t}^{se}])=\mathcal{A}[\rho _{t}^{e}],
\label{Condition}
\end{equation}%
where $\mathcal{A}$ is an arbitrary superoperator acting on $\rho _{t}^{e}.$
This criterion allows us to find which kind of system-environment couplings
fulfill the proposed definition.

\subsection{Unitary coupling}

A unitary coupling is set by a bipartite Hamiltonian $H_{se}$ such that%
\begin{equation}
\mathcal{L}_{se}[\bullet ]=-i[H_{se},\bullet ].
\end{equation}%
In order to check condition (\ref{Condition}), we introduce a complete
orthogonal basis $\{|s\rangle \}$ of the system Hilbert space such that $%
\sum_{s}|s\rangle \langle s|=\mathrm{I}_{s},$ where $\mathrm{I}_{s}$ is the
system identity matrix. We get,%
\begin{equation}
\mathrm{Tr}_{s}(\mathcal{L}_{se}[\rho _{t}^{se}])=-i\sum_{s,s^{\prime }}%
\left[ \langle s|H_{se}|s^{\prime }\rangle ,\langle s^{\prime }|\rho
_{t}^{se}|s\rangle \right] .
\end{equation}%
Thus, independence of the environment state of the system degrees of freedom
requires $\langle s|H_{se}|s^{\prime }\rangle =\delta _{s,s^{\prime }}Q_{e},$
which in turn implies $H_{se}=I_{s}\otimes Q_{e},$ where $Q_{e}=Q_{e}^{\dag
} $ is an arbitrary operator acting on $\mathcal{H}_{e}.$ Nevertheless, this
solution implies that system and environment do not interact. Consequently,
it is impossible to obtain a casual bystander \textit{non-Markovian}
environment if it interacts unitarily with the system of interest.

The previous result is not valid when a Born-Markov approximation applies 
\cite{breuerbook}, where the environment state can be approximated by a
stationary one, $\rho _{t}^{e}=\rho _{0}^{e}.$ Nevertheless, in this
situation memory effects do not develop.

\subsection{Dissipative coupling}

After discarding the unitary property, now we consider dissipative
system-environment couplings. Thus, the environment is defined by a set of
quantum degrees of freedom whose interaction with the system can be
approximated by an arbitrary (non-diagonal) Lindblad superoperator \cite%
{breuerbook},%
\begin{equation}
\mathcal{L}_{se}[\bullet ]=\sum_{i,j}\gamma _{i,j}(T_{i}\bullet T_{j}^{\dag
}-\frac{1}{2}\{T_{j}^{\dag }T_{i},\bullet \}_{+}),
\end{equation}%
where $\{a,b\}_{+}\equiv ab+ba.$ The complex parameters $\{\gamma _{i,j}\}$
define the (diagonal and nondiagonal) rate coefficients of the dissipative
channels corresponding to the bipartite operators $\{T_{i}\}.$ Thus, they
constitute an Hermitian positive definite matrix \cite{alicki}. Under the
replacements $T_{i}\rightarrow V_{k}\otimes B_{\alpha },\ T_{j}\rightarrow
V_{l}\otimes B_{\beta }$\ and $\gamma _{i,j}\rightarrow \gamma _{k\alpha
,l\beta },$ where $\{V_{k}\}$ and $\{B_{\alpha }\}$ are arbitrary operators
acting on $\mathcal{H}_{s}$ and $\mathcal{H}_{e}$ respectively, $\mathcal{L}%
_{se}$ can be rewritten as%
\begin{equation}
\mathcal{L}_{se}[\bullet ]\!=\!\!\sum_{k\alpha ,l\beta }\gamma _{k\alpha
,l\beta }(V_{k}B_{\alpha }\bullet B_{\beta }^{\dag }V_{l}^{\dag }-\frac{1}{2}%
\{V_{l}^{\dag }V_{k}B_{\beta }^{\dag }B_{\alpha },\bullet \}_{+}).
\label{BipartiteGenerator}
\end{equation}%
Here, the indexes run in the intervals $k,l=1,\cdots ,(\dim \mathcal{H}%
_{s})^{2}-1,$ and $\alpha ,\beta =1,\cdots ,(\dim \mathcal{H}_{e})^{2}-1.$

By taking the trace to the interaction superoperator $\mathcal{L}_{se}[\rho
_{t}^{se}],$ we get%
\begin{eqnarray*}
\mathrm{Tr}_{s}(\mathcal{L}_{se}[\rho _{t}^{se}])\! &=&\!\!\sum_{\alpha
,\beta ,s}\Big{(}B_{\alpha }\langle s|D_{\alpha ,\beta }\rho
_{t}^{se}|s\rangle B_{\beta }^{\dag }-\frac{1}{2}B_{\beta }^{\dag }B_{\alpha
} \\
&&\!\!\!\!\times \langle s|D_{\alpha ,\beta }\rho _{t}^{se}|s\rangle -\frac{1%
}{2}\langle s|\rho _{t}^{se}D_{\alpha ,\beta }|s\rangle B_{\beta }^{\dag
}B_{\alpha }\Big{)},
\end{eqnarray*}%
where as before $\{|s\rangle \}$ is a complete base in $\mathcal{H}_{s}.$
Furthermore, we introduced the system operators $\{D_{\alpha ,\beta }\},$%
\begin{equation}
D_{\alpha ,\beta }\equiv \sum_{k,l}\gamma _{k\alpha ,l\beta }V_{l}^{\dag
}V_{k},\ \ \ \ \ \ \ \ D_{\alpha ,\beta }^{\dag }=D_{\beta ,\alpha }.
\end{equation}%
The Hermitian property is inherited from condition\ $\rho _{t}^{se}=(\rho
_{t}^{se})^{\dagger }$ in Eq.~(\ref{RhoSUEvolution}). It is simple to
realize that the constraint~(\ref{Condition}) is fulfilled if%
\begin{equation}
\langle s|D_{\alpha ,\beta }=\Gamma _{\alpha ,\beta }\langle s|,\ \ \ \ \ \
\ \ D_{\alpha ,\beta }|s\rangle =\Gamma _{\alpha ,\beta }|s\rangle ,
\end{equation}%
where $\Gamma _{\alpha ,\beta }$ are, in general, complex coefficients.
Applying $\sum_{s^{\prime }}|s^{\prime }\rangle $ and $\sum_{s^{\prime
}}\langle s^{\prime }|$ to the left and right equalities, using that $%
D_{\alpha ,\beta }^{\dag }=D_{\beta ,\alpha },$ we get the equivalent
conditions%
\begin{equation}
D_{\alpha ,\beta }=\Gamma _{\alpha ,\beta }\mathrm{I}_{s},\ \ \ \ \ \ \ \
\Gamma _{\alpha ,\beta }^{\ast }=\Gamma _{\beta ,\alpha }.
\label{FinalConstraint}
\end{equation}

Introducing the final constraints~(\ref{FinalConstraint}) into Eq.~(\ref%
{BipartiteGenerator}), we can write the bipartite interaction generator as%
\begin{equation}
\mathcal{L}_{se}[\bullet ]=\sum_{\alpha ,\beta }\Gamma _{\alpha ,\beta }\
(B_{\alpha }\mathbb{S}_{\alpha ,\beta }[\bullet ]B_{\beta }^{\dag }-\frac{1}{%
2}\{B_{\beta }^{\dag }B_{\alpha },\bullet \}_{+}),
\end{equation}%
where the coefficients $\{\Gamma _{\alpha ,\beta }\}$ define an Hermitian
positive definite matrix. Furthermore, $\{\mathbb{S}_{\alpha ,\beta }\}$ is
a set of \textit{arbitrary} completely positive system superoperators that
fulfill the symmetry $\mathbb{S}_{\alpha ,\beta }^{\dag }=\mathbb{S}_{\beta
,\alpha }$ and are trace preserving, $\mathrm{Tr}_{s}[\mathbb{S}_{\alpha
,\beta }[\rho ^{s}]]=\mathrm{Tr}_{s}[\rho ^{s}].$ Hence, they can be written
in a Kraus representation \cite{breuerbook} as%
\begin{equation}
\mathbb{S}_{\alpha ,\beta }[\bullet ]=\sum_{k}V_{k}^{\alpha \beta }\bullet
V_{k}^{\dag \alpha \beta },\ \ \ \ \ \ \ \ \ \ \ \ \ \sum_{k}V_{k}^{\dag
\alpha \beta }V_{k}^{\alpha \beta }=\mathrm{I}_{s},  \label{Kraus}
\end{equation}%
where $V_{k}^{\alpha \beta }$ are (arbitrary) system operators. The
bipartite system-environment evolution [Eq.~(\ref{RhoSUEvolution})] can
finally be written as%
\begin{eqnarray}
\frac{d}{dt}\rho _{t}^{se} &=&(\mathcal{L}_{s}+\mathcal{L}_{e})[\rho
_{t}^{se}]+\sum_{\alpha ,\beta }\Gamma _{\alpha ,\beta }\ B_{\alpha }\mathbb{%
S}_{\alpha ,\beta }[\rho _{t}^{se}]B_{\beta }^{\dag }  \notag \\
&&-\frac{1}{2}\sum_{\alpha ,\beta }\Gamma _{\alpha ,\beta }\{B_{\beta
}^{\dag }B_{\alpha },\rho _{t}^{se}\}_{+},  \label{GeneralSU}
\end{eqnarray}%
where $\mathcal{L}_{s}$ and $\mathcal{L}_{e}$ are arbitrary. This equation
is the main result of this section. It defines the most general dissipative
system-environment coupling that is consistent with a quantum non-Markovian
casual bystander environment. In fact, after applying the (system) trace
operation to Eq.~(\ref{GeneralSU}), and using the property~(\ref{Kraus}),
the density matrix $\rho _{t}^{e}$ of the environment [Eq.~(\ref{Rhoes})]
evolves as%
\begin{equation}
\frac{d}{dt}\rho _{t}^{e}=\mathcal{L}_{e}[\rho _{t}^{e}]+\sum_{\alpha ,\beta
}\Gamma _{\alpha ,\beta }\ (B_{\alpha }\rho _{t}^{e}B_{\beta }^{\dag }-\frac{%
1}{2}\{B_{\beta }^{\dag }B_{\alpha },\rho _{t}^{e}\}_{+}).  \label{GeneralU}
\end{equation}%
As expected, this Lindblad equation does not depend on the system degrees of
freedom. On the other hand, the time evolution of the system state, $\rho
_{t}^{s}=\mathrm{Tr}_{e}[\rho _{t}^{se}],$ assuming uncorrelated initial
conditions $\rho _{0}^{se}=\rho _{0}^{s}\otimes \rho _{0}^{e},$ from Eq.~(%
\ref{GeneralSU}) can formally be written as a time-convoluted equation,%
\begin{equation}
\frac{d}{dt}\rho _{t}^{s}=\mathcal{L}_{s}[\rho
_{t}^{s}]+\int_{0}^{t}dt^{\prime }\mathcal{K}_{s}(t-t^{\prime })[\rho
_{t^{\prime }}^{s}],  \label{SystemConvolution}
\end{equation}%
where the superoperator $\mathcal{K}_{s}(t)$ is defined in a Laplace domain $%
[f(z)=\int_{0}^{\infty }dte^{-zt}f(t)]$ from the relation $\mathrm{Tr}_{e}[%
\mathcal{G}_{z}^{se}(\mathcal{L}_{e}+\mathcal{L}_{se})\rho _{0}^{e}][\bullet
]=\mathrm{Tr}_{e}[\mathcal{G}_{z}^{se}\rho _{0}^{e}]\mathcal{K}%
_{s}(z)[\bullet ],$ where $\mathcal{G}_{z}^{se}=[z-(\mathcal{L}_{s}+\mathcal{%
L}_{e}+\mathcal{L}_{se})]^{-1}$ is the bipartite system-environment
propagator.

Both, Eqs.~(\ref{GeneralSU}) and (\ref{GeneralU}) can always be reduced to a
standard diagonal form~\cite{breuerbook,alicki}, which can be read by taking 
$\Gamma _{\alpha ,\beta }=\delta _{\alpha ,\beta }\Gamma _{\alpha },$ where $%
\{\Gamma _{\alpha }\}$ are positive rate coefficients. Furthermore, Eq.~(\ref%
{SystemConvolution}) can always be transformed into a convolutionless form~%
\cite{LocalNonLocal}.

\section{General properties}

On the basis of Eq.~(\ref{GeneralSU}), it is possible to establish general
properties that characterize the system-environment dynamics.

\subsection{System-environment correlations}

Even for uncorrelated initial conditions, the evolution~(\ref{GeneralSU})
induces correlations between the system and the degrees of freedom
associated to the environment. These correlations can be characterized from
the bipartite state $\rho _{t}^{se}.$ Here, we show that, for uncorrelated
initial conditions, $\rho _{0}^{se}=\rho _{0}^{s}\otimes \rho _{0}^{e},$ it
always assumes the structure%
\begin{equation}
\rho _{t}^{se}=\sum_{c}\rho _{c}(t)\otimes |c_{t}\rangle \langle c_{t}|,
\label{FormalSolution}
\end{equation}%
where $\{\rho _{c}(t)\}$ are states (matrixes)\ in $\mathcal{H}_{s}$ and $%
\{|c_{t}\rangle \langle c_{t}|\}$ are orthogonal time-dependent projectors
in $\mathcal{H}_{e},$ that is, $\langle c_{t}|c_{t}^{\prime }\rangle =\delta
_{cc^{\prime }}.\ $Consistently, the system and environment states read%
\begin{equation}
\rho _{t}^{s}=\sum_{c}\rho _{c}(t),\ \ \ \ \ \ \ \ \ \rho
_{t}^{e}=\sum_{c}p_{c}(t)|c_{t}\rangle \langle c_{t}|,  \label{RhosSysEnv}
\end{equation}%
where $p_{c}(t)\equiv \mathrm{Tr}_{s}[\rho _{c}(t)].$ From these
expressions, it follows that $\{|c_{t}\rangle \}$ is the base in which $\rho
_{t}^{e}$ becomes diagonal at time $t.$ Furthermore, $\rho _{c}(t)$ is the
conditional state of the system given that the environment is in the state $%
|c_{t}\rangle \langle c_{t}|.$

The formal solution~(\ref{FormalSolution}) implies that $\rho _{t}^{se}$ is
a \textit{separable state} \cite{entanglement} with a null
system-environment discord \cite{discord}. Hence, not any quantum
entanglement (between the system and the environment) is produced during the
evolution.

The validity of Eq.~(\ref{FormalSolution}) can be established from the
bipartite evolution~(\ref{GeneralSU}). From these equations, for the system
conditional states $\{\rho _{c}(t)\},$ we get the evolution%
\begin{eqnarray}
\frac{D}{Dt}\rho _{c}(t) &=&\mathcal{L}_{s}[\rho _{c}(t)]+\sum_{\tilde{c}%
}\phi _{c\tilde{c}}(t)\rho _{\tilde{c}}(t)-\sum_{\tilde{c}}\phi _{\tilde{c}%
c}(t)\rho _{c}(t)  \notag \\
&&\!\!\!\!\!\!+\sum_{\tilde{c}}\gamma _{c\tilde{c}}(t)\mathbb{S}_{c\tilde{c}%
}(t)[\rho _{\tilde{c}}(t)]-\sum_{\tilde{c}}\gamma _{\tilde{c}c}(t)\rho
_{c}(t).  \label{DRhoC}
\end{eqnarray}%
Here, a \textquotedblleft total-time-derivative\textquotedblright\ was
introduced,%
\begin{equation}
\frac{D}{Dt}\rho _{c}(t)\equiv \frac{d}{dt}\rho _{c}(t)+\sum_{\tilde{c}%
}\langle c_{t}|\frac{d}{dt}[\Pi _{t}^{\tilde{c}}]|c_{t}\rangle \rho _{\tilde{%
c}}(t),  \label{Dtotal}
\end{equation}%
where $\Pi _{t}^{c}\equiv |c_{t}\rangle \langle c_{t}|.$ Furthermore, the
time-dependent rates are%
\begin{equation}
\phi _{c\tilde{c}}(t)=\langle c_{t}|\mathcal{L}_{e}[\Pi _{t}^{\tilde{c}%
}]|c_{t}\rangle \geq 0,
\end{equation}%
and similarly%
\begin{equation}
\gamma _{\tilde{c}c}(t)=\sum_{\alpha ,\beta }\Gamma _{\alpha ,\beta }\
\langle c_{t}|B_{\alpha }|\tilde{c}_{t}\rangle \langle \tilde{c}%
_{t}|B_{\beta }^{\dag }|c_{t}\rangle \geq 0,
\end{equation}%
where the inequality follows straightforwardly from the diagonal rate
representation $\Gamma _{\alpha ,\beta }=\delta _{\alpha ,\beta }\Gamma
_{\alpha }.$ Finally, in Eq.~(\ref{DRhoC}) the system superoperators $%
\mathbb{S}_{c\tilde{c}}(t)[\bullet ]$ read%
\begin{equation}
\mathbb{S}_{c\tilde{c}}(t)[\bullet ]=\frac{\sum_{\alpha ,\beta }\Gamma
_{\alpha ,\beta }\ \langle c_{t}|B_{\alpha }|\tilde{c}_{t}\rangle \langle 
\tilde{c}_{t}|B_{\beta }^{\dag }|c_{t}\rangle \mathbb{S}_{\alpha ,\beta
}[\bullet ]}{\sum_{\alpha ,\beta }\Gamma _{\alpha ,\beta }\ \langle
c_{t}|B_{\alpha }|\tilde{c}_{t}\rangle \langle \tilde{c}_{t}|B_{\beta
}^{\dag }|c_{t}\rangle },
\end{equation}%
which are trace preserving, $\mathrm{Tr}_{s}[\mathbb{S}_{c\tilde{c}}(t)[\rho
]]=\mathrm{Tr}_{s}[\rho ].$ The system superoperators $\mathbb{S}_{\alpha
,\beta }[\bullet ]$\ are defined by Eq.~(\ref{Kraus}).

On the other hand, the evolution of the probabilities $\{p_{c}(t)\}$ [Eq.~(%
\ref{RhosSysEnv})] follows by taking the trace of Eq.~(\ref{DRhoC}), which
yields%
\begin{eqnarray}
\frac{D}{Dt}p_{c}(t) &=&+\sum_{\tilde{c}}\phi _{c\tilde{c}}(t)p_{\tilde{c}%
}(t)-\sum_{\tilde{c}}\phi _{\tilde{c}c}(t)p_{c}(t)  \notag \\
&&\!\!+\sum_{\tilde{c}}\gamma _{c\tilde{c}}(t)p_{\tilde{c}}(t)-\sum_{\tilde{c%
}}\gamma _{\tilde{c}c}(t)p_{c}(t).  \label{Dpc}
\end{eqnarray}%
Here, $(D/Dt)p_{c}(t)$ follows from Eq.~(\ref{Dtotal}) under the replacement 
$\rho _{c}(t)\rightarrow p_{c}(t).$ Consistently, this equation does not
depend on the system degrees of freedom.

The consistence of the time evolution of $\rho _{c}(t)$ and $p_{c}(t)$
[Eqs.~(\ref{DRhoC}) and (\ref{Dpc}) respectively] supports the structure
defined by Eq.~(\ref{FormalSolution}). The physical meaning of these results
can be easily understood by analyzing an incoherent environment case.

\subsection{Incoherent environment}

If the degrees of freedom of the environment do not develop any (quantum)
coherence, at any time its density matrix $\rho _{t}^{e}$\ is diagonal in a
fixed base $\{|c\rangle \}.$ Thus, the previous results must be read under
the replacement%
\begin{equation}
|c_{t}\rangle \langle c_{t}|\rightarrow |c\rangle \langle c|.
\label{IncoherentCond}
\end{equation}%
In consequence, the total-time-derivative [Eq.~(\ref{Dtotal})] becomes an
usual time-derivative, $(D/Dt)\rightarrow (d/dt),$ and the rates $\phi _{c%
\tilde{c}}(t)$ and $\gamma _{\tilde{c}c}(t)$ do not depend on time. In this
situation, the probabilities $\{p_{c}(t)\}$ obey a standard classical master
equation [Eq.~(\ref{Dpc})]. Similarly, the evolution of the states $\{\rho
_{c}(t)\}$ [Eq.~(\ref{DRhoC})]\ involves the same classical coupling with
rates $\phi _{c\tilde{c}},$ while the coupling with rates $\gamma _{c\tilde{c%
}}$ are endowed by the application of the superoperators $\mathbb{S}_{c%
\tilde{c}}$ in each environment transition $c\leftarrow \tilde{c}.$ Thus,
the evolution becomes a particular case corresponding to a system driven by
incoherent degrees of freedom that follows their own stochastic dynamics
(see Eq.~(27) in Ref.~\cite{maximal}).

When condition~(\ref{IncoherentCond}) is not fulfilled, the interpretation
of Eqs.~(\ref{DRhoC}) and (\ref{Dpc}) remains the same as in the incoherent
case. Nevertheless, the base $\{|c_{t}\rangle \}$ associated to the
environment become time-dependent due to the intrinsic quantum nature of the
environment degrees of freedom. This effect is taken into account through
the total-time-derivative $D/Dt$ [Eq.~(\ref{Dtotal})].

\subsection{Measurement based stochastic representation}

A clear understanding of the system-environment coupling can be achieved by
representing their dynamics with a measurement-based~\cite%
{carmichaelbook,plenio} stochastic bipartite state $\rho _{st}^{se}(t).$
Thus, in average over realizations (denoted with an overbar symbol) it
follows%
\begin{equation}
\rho _{t}^{se}=\overline{\rho _{st}^{se}(t)}.
\end{equation}%
In principle, the state $\rho _{st}^{se}(t)$ can be formulated from the
incoherent-like representation~(\ref{DRhoC}). A deeper understanding is
achieved by assuming that the degrees of freedom of the environment are
subjected to continuous-in-time measurement process that resolves the
transitions induced by the (bath) operators $\{B_{\alpha }\}.$ Thus, $\rho
_{st}^{se}(t)$ follows from the standard quantum jump approach~\cite%
{carmichaelbook,plenio} (for simplicity we consider in Eqs.~(\ref{GeneralSU}%
) and (\ref{GeneralU}) the diagonal case $\Gamma _{\alpha ,\beta }=\delta
_{\alpha ,\beta }\Gamma _{\alpha },$ denoting $\mathbb{S}_{\alpha ,\alpha
}\leftrightarrow \mathbb{S}_{\alpha }).$ The environment state is recovered
as $\rho _{t}^{e}=\overline{\rho _{st}^{e}(t)},$ where $\rho _{st}^{e}(t)=%
\mathrm{Tr}_{s}[\rho _{st}^{e}(t)].$ For a casual bystander environment, the
evolution of $\rho _{st}^{e}(t)$ is independent of the system dynamics,
being defined by the transitions associated to $\{B_{\alpha }\}.$ Similarly,
from the evolution~(\ref{GeneralSU}) it follows that the bipartite state
(assuming uncorrelated initial conditions), must to take the form%
\begin{equation}
\rho _{st}^{sa}(t)=\rho _{st}^{s}(t)\otimes \rho _{st}^{e}(t).
\label{StochasticRep}
\end{equation}%
It is simple to realize that the dynamics of the system state $\rho
_{st}^{s}(t)$ must include the action of the superoperator $\mathbb{S}%
_{\alpha }$ whenever the environment suffers a transition (jump)
corresponding to the operator $B_{\alpha }.$ In this way, the bipartite
system-environment correlations [Eq.~(\ref{FormalSolution})] are built up in
average. On the other hand, between environment transitions, the state $\rho
_{st}^{s}(t)$ evolves under the action of $\mathcal{L}_{s}.$ Thus, the
system dynamics can be seen as a \textit{collisional} one \cite%
{collisionVacchini,embedding}, where the occurrence of the sudden
(collisional) changes $\rho _{st}^{s}\rightarrow \mathbb{S}_{\alpha }[\rho
_{st}^{s}]$ are dictated by the environment transitions associated to the
operator $B_{\alpha }.$ We notice that this representation recover the
results of Ref.~\cite{embedding}, which can be read as a particular case of
the general dynamics (\ref{GeneralSU}).

\subsection{Quantum regression hypothesis}

The underlying system-environment dynamics is Markovian [Eq.~(\ref{GeneralSU}%
)]. Thus, the quantum regression theorem (QRT)~\cite{carmichaelbook} is
valid in the bipartite space $\mathcal{H}_{s}\otimes \mathcal{H}_{e}.$
Introducing a vector of system operators $\mathbf{A\leftrightarrow A}\otimes 
\mathrm{I}_{e},$ where $\mathbf{A}=(A_{1},A_{2,}\cdots ,A_{\dim (\mathcal{H}%
_{s})^{2}}),$ their expectation value at a time $\tau $ (for simplicity,
also denoted with an overbar symbol) can be written as%
\begin{equation}
\overline{\mathbf{A}(\tau )}=\mathrm{Tr}_{se}(\mathcal{G}_{\tau
,0}^{se}[\rho _{0}^{se}]\mathbf{A}).  \label{OperatorBipartite}
\end{equation}%
The bipartite propagator is $\mathcal{G}_{\tau ,\tau _{0}}^{se}\equiv \exp
[(\tau -\tau _{0})\mathcal{L}_{T}],$ with $\mathcal{L}_{T}\equiv (\mathcal{L}%
_{s}+\mathcal{L}_{e}+\mathcal{L}_{se}),$ where $\mathcal{L}_{se}$ follows
from Eq.~(\ref{GeneralSU}). Given an extra system operator $O\leftrightarrow
O\otimes \mathrm{I}_{e},$ the correlations $\overline{O(t)\mathbf{A}(t+\tau )%
}$ follow from the QRT, which implies~\cite{carmichaelbook}%
\begin{equation}
\overline{O(t)\mathbf{A}(t+\tau )}=\mathrm{Tr}_{se}(\mathbf{A}\mathcal{G}%
_{t+\tau ,t}^{se}[\rho _{t}^{se}O\mathbb{]}).  \label{Correlations}
\end{equation}

Now, we search conditions under which the QRT is valid on $\mathcal{H}_{s}.$
Thus, we explore if the previous (system) operator correlations can be
written only in terms of the system propagator. Assuming uncorrelated
initial conditions, $\rho _{0}^{se}=\rho _{0}^{s}\otimes \rho _{0}^{e},$ the
system propagator $\mathcal{G}_{\tau ,0}^{s}$\ can be written as%
\begin{equation}
\mathcal{G}_{\tau ,0}^{s}[\bullet ]\equiv \mathrm{Tr}_{e}(\mathcal{G}_{\tau
,0}^{se}[\bullet ]\otimes \rho _{0}^{e}]).
\end{equation}%
Thus, from Eq.~(\ref{OperatorBipartite}), the operator expectation values
can be rewritten as%
\begin{equation}
\overline{\mathbf{A}(\tau )}=\mathrm{Tr}_{s}(\mathcal{G}_{\tau ,0}^{s}[\rho
_{0}^{s}]\mathbf{A}).
\end{equation}%
On the other hand, it is simple to realize that the correlations~(\ref%
{Correlations}) cannot be written only in terms of the system propagator $%
\mathcal{G}_{t+\tau ,t}^{s},$ which implies that the QRT is not valid in
general on $\mathcal{H}_{s}.$ Nevertheless,\textit{\ assuming} that the
environment begins in its stationary state $(\rho _{\mathrm{\infty }}^{e}),$
and that the stationary bipartite state $(\lim\nolimits_{t\rightarrow \infty
}\rho _{t}^{se})$ do not involve system-environment correlations,%
\begin{equation}
\rho _{0}^{se}=\rho _{0}^{s}\otimes \rho _{\mathrm{\infty }}^{e},\ \ \ \ \ \
\ \ \ \ \ \ \lim\nolimits_{t\rightarrow \infty }\rho _{t}^{se}=\rho _{%
\mathrm{\infty }}^{s}\otimes \rho _{\mathrm{\infty }}^{e},
\label{QRTConditions}
\end{equation}%
in the\textit{\ long time limit }the operator correlations become%
\begin{equation}
\lim\nolimits_{t\rightarrow \infty }\overline{O(t)\mathbf{A}(t+\tau )}=%
\mathrm{Tr}_{s}(\mathbf{A}\mathcal{G}_{\tau ,0}^{s}[\rho _{\mathrm{\infty }%
}^{s}O\mathbb{]}).
\end{equation}%
Thus, if the conditions~(\ref{QRTConditions}) are fulfilled the QRT is valid
in the stationary regime \textit{even} when the system dynamics is
non-Markovian. Interestingly, the same property arises in\ quantum systems
coupled to arbitrary incoherent degrees of freedom~\cite{QRTLindbladRate}.
In fact, given that none condition on $\mathcal{G}_{\tau ,0}^{se}$ was
demanded in the previous derivation, this result is valid in general
whenever the bipartite (system-environment) dynamics is a Markovian
(Lindblad) one. The explicit meaning of the restricted validity of the QRT
becomes clear by writing $\overline{\mathbf{A}(\tau )}=\boldsymbol{\hat{T}}%
(\tau )\overline{\mathbf{A}(0)},$ and the stationary correlations as $%
\lim\nolimits_{t\rightarrow \infty }\overline{O(t)\mathbf{A}(t+\tau )}=%
\boldsymbol{\hat{T}}(\tau )\lim\nolimits_{t\rightarrow \infty }\overline{O(t)%
\mathbf{A}(t)},$ where $\boldsymbol{\hat{T}}(\tau )$ is a matrix in the
space corresponding to the vector of operators $\overline{\mathbf{A}(0)}$~%
\cite{carmichaelbook}.

\subsection{Operational memory witness}

Memory effects can alternatively be characterized from measurement-based
approaches. A conditional past-future (CPF) correlation \cite{budiniCPF} is
defined by a set of three successive measurements performed over the system
of interest. Here, they are taken as projective ones, corresponding to
Hermitian operators $O_{\underline{\mathbf{m}}},$ denoted in successive
order with$\ \underline{\mathbf{m}}=\underline{\mathbf{x}},\underline{%
\mathbf{y}},\underline{\mathbf{z}}.$ Their eigenvectors and eigenvalues read 
$O_{\underline{\mathbf{m}}}|m\rangle =m|m\rangle ,\ $where correspondingly $%
\{m\}=\{x\},$ $\{y\},$ $\{z\}.$ The measurements are performed at the
initial time $t=0$ (past), at time $t$ (present) and $t+\tau $ (future)
respectively. After the intermediate measurement at time $t,$ the system
post-measurement state is externally modified~\cite{BIF} as $\rho
_{y}=|y\rangle \langle y|\rightarrow \rho _{\breve{y}}=|\breve{y}\rangle
\langle \breve{y}|.$ A \textit{deterministic scheme} (d) is defined by the
condition $\breve{y}=y.$ Thus, not any change is introduced. A \textit{%
random scheme} (r) is defined by a random election of $\breve{y}$ (over the
set $\{y\}$) with and arbitrary probability $\wp (\breve{y}|x),$ which may
depend on the outcomes $\{x\}$ of the first measurement performed at time $%
t=0.$ The CPF correlation depends on the chosen scheme. In both cases, it
reads%
\begin{equation}
C_{pf}(t,\tau )|_{\breve{y}}\overset{d/r}{=}\sum_{z,x}zx[P(z,x|\breve{y}%
)-P(z|\breve{y})P(x|\breve{y})],  \label{CPFexplicit}
\end{equation}%
where $\{z\}$ and $\{x\}$ are the eigenvalues of $O_{\underline{\mathbf{z}}}$
and $O_{\underline{\mathbf{x}}}$ respectively. With $P(a|b)$ we denote the
conditional probability of $a$ given $b.$

In the deterministic scheme, the CPF correlation vanishes in a Markovian
regime, where past and future outcomes are conditionally independent: $P(z,x|%
\breve{y})=P(z|\breve{y})P(x|\breve{y})$ \cite{budiniCPF}. Thus, $%
C_{pf}(t,\tau )|_{\breve{y}}\overset{d}{\neq }0$ detects memory effects
independently of their underling mechanism. This last property can be
understood from the change of the bipartite state $\rho _{se}$ after the
intermediate (projective) measurement,%
\begin{equation}
\rho _{se}\rightarrow |y\rangle \langle y|\otimes \frac{\langle y|\rho
_{se}|y\rangle }{\mathrm{Tr}_{se}[|y\rangle \langle y|\rho _{se}]}.
\label{RhoTransform}
\end{equation}%
The dependence of the environment post-measurement state $\langle y|\rho
_{se}|y\rangle /\mathrm{Tr}_{se}[|y\rangle \langle y|\rho _{se}]$ on the
previous system history (measurement outcomes at the initial time) allows to
detect memory effects.

In the random scheme, even in presence of memory effects, the CPF
correlation vanishes when the environment is not affected during the system
evolution~\cite{BIF}. Consequently, the presence of \textit{a
(non-Markovian) casual bystander environment implies that} $C_{pf}(t,\tau
)|_{\breve{y}}\overset{r}{=}0.$ Alternatively, in the random scheme, the
condition $C_{pf}(t,\tau )|_{\breve{y}}\overset{r}{\neq }0$ indicates
departures with respect to a casual bystander environment, here defined by
the condition Eq.~(\ref{Condition}).

The previous features of the random scheme can be understood from Eq.~(\ref%
{RhoTransform}) by introducing the random system transformation $|y\rangle
\langle y|\rightarrow |\breve{y}\rangle \langle \breve{y}|$ and averaging
(marginating) the environment post-measurement states (associated to the
outcomes $\{y\}$) with their probabilities $\{\mathrm{Tr}_{se}[|y\rangle
\langle y|\rho _{se}]\}.$ Thus, after the intermediate measurement the
bipartite state transform as%
\begin{equation}
\rho _{se}\rightarrow |\breve{y}\rangle \langle \breve{y}|\otimes
\sum_{y}\langle y|\rho _{se}|y\rangle =|\breve{y}\rangle \langle \breve{y}%
|\otimes \mathrm{Tr}_{s}[\rho _{se}].
\end{equation}%
From the definition of a casual bystander environment [see also Eq.~(\ref%
{Condition})], it follows that the (bath) state $\mathrm{Tr}_{s}[\rho _{se}]$
does not depend on the previous (system) history, which implies that a
Markov property characterize the outcome statistics. Consequently, the CPF
correlation vanishes.

The conditional probabilities appearing in Eq.~(\ref{CPFexplicit}) can
explicitly be calculated from the joint outcome probability $P(z,\check{y}%
,x)\leftrightarrow P(z,t+\tau ;\check{y},t;x,0).$ As shown in Refs.~\cite%
{budiniCPF,BIF}, this object can be calculated after knowing the
system-environment propagator. In order to maintain a description as simple
as possible, we consider Eq.~(\ref{GeneralSU}) in the diagonal case $\Gamma
_{\alpha ,\beta }=\delta _{\alpha ,\beta }\Gamma _{\alpha }$ (denoting $%
\mathbb{S}_{\alpha ,\alpha }\leftrightarrow \mathbb{S}_{\alpha }$) and
assuming that the composition of two arbitrary superoperators can be written
as a superoperator included in the master equation, that is, $\mathbb{S}%
_{\alpha }\mathbb{S}_{\alpha ^{\prime }}=\mathbb{S}_{\alpha ^{\prime \prime
}}.$ Under these two conditions, the bipartite propagator, $\rho _{t}^{se}=%
\mathcal{G}_{t,0}^{se}[\rho _{0}^{s}\otimes \rho _{0}^{e}],$ can be written
in a general way as%
\begin{equation}
\rho _{t}^{se}=\rho _{0}^{s}\otimes \mathcal{F}_{0}(t)[\rho _{0}^{e}]+\sum 
_{\substack{ \alpha  \\ \alpha \neq 0}}\mathbb{S}_{\alpha }[\rho
_{0}^{s}]\otimes \mathcal{F}_{\alpha }(t)[\rho _{0}^{e}].
\label{SEPropagator}
\end{equation}%
Here, the environment superoperators $\{\mathcal{F}_{\alpha }(t)\},$ which
act on the initial environment state $\rho _{0}^{e},$ depend on each
specific problem.

\subsubsection{Deterministic scheme}

In the deterministic scheme, the joint outcome probability can be written as~%
\cite{BIF}%
\begin{equation}
\frac{P(z,\breve{y},x)}{P(x)}\overset{d}{=}\mathrm{Tr}_{se}(E_{z}\mathcal{G}%
_{t+\tau ,t}^{se}[\rho _{\breve{y}}\otimes \mathrm{Tr}_{s}(E_{\breve{y}}%
\mathcal{G}_{t,0}^{se}[\rho _{x}^{se}])]).
\end{equation}%
Here, $E_{m}\equiv |m\rangle \langle m|,$ $\rho _{m}\equiv |m\rangle \langle
m|$ $[m=z,\breve{y},x]$ and $\rho _{x}^{se}\equiv \rho _{x}\otimes \rho
_{0}^{e}.$ Furthermore, $P(x)=\langle x|\rho _{0}^{s}|x\rangle .$ From this
expression, using the bipartite propagator~(\ref{SEPropagator}), we get%
\begin{eqnarray}
&&\frac{P(z,\check{y},x)}{P(x)}\overset{d}{=}\sum_{\alpha ,\beta }\langle z|%
\mathbb{S}_{\alpha }[\rho _{\check{y}}]|z\rangle \langle \check{y}|\mathbb{S}%
_{\beta }[\rho _{x}]|\check{y}\rangle  \notag \\
&&\ \ \ \ \ \ \ \ \ \ \ \ \ \ \ \ \ \times \mathrm{Tr}_{e}[\mathcal{F}%
_{\alpha }(\tau )\mathcal{F}_{\beta }(t)\rho _{0}^{e}\mathbb{].}
\label{P3Deter}
\end{eqnarray}%
Using that $P(z,x|\breve{y})=P(z,\breve{y},x)/P(\breve{y}),$ where $P(\breve{%
y})=\sum_{z,x}P(z,\breve{y},x),$ the CPF correlation~(\ref{CPFexplicit})
reads%
\begin{equation}
C_{pf}(t,\tau )|_{\breve{y}}\overset{d}{=}\frac{1}{P(\breve{y})^{2}}%
\sum_{\alpha ,\beta ,\mu }\Theta ^{\alpha \beta \mu }|_{\breve{y}}\Lambda
_{\alpha \beta \mu }(t,\tau ).  \label{CPFGen}
\end{equation}%
The time-independent coefficients $\Theta ^{\alpha \beta \mu }|_{y}$ only
depend on the chosen observables,%
\begin{equation}
\Theta ^{\alpha \beta \mu }|_{\breve{y}}=\langle \breve{y}|\mathbb{S}%
_{\alpha }^{\#}[O_{\underline{\mathbf{z}}}]|\breve{y}\rangle \langle \breve{y%
}|\mathbb{S}_{\beta }[O_{\underline{\mathbf{x}}}\rho _{\underline{\mathbf{x}}%
}]|\breve{y}\rangle \langle \breve{y}|\mathbb{S}_{\mu }[\rho _{\underline{%
\mathbf{x}}}]|\breve{y}\rangle ,
\end{equation}%
where the dual superoperator $\mathbb{S}_{\alpha }^{\#}$ is defined from $%
\mathrm{Tr}_{s}(O\mathbb{S}_{\alpha }[\rho ])=\mathrm{Tr}_{s}(\rho \mathbb{S}%
_{\alpha }^{\#}[O]).$ Furthermore, we defined the state $\rho _{\underline{%
\mathbf{x}}}\equiv \sum_{x}P(x)|x\rangle \langle x|=\sum_{x}\langle x|\rho
_{0}^{s}|x\rangle |x\rangle \langle x|.$ The time-dependence in Eq.~(\ref%
{CPFGen}) follows from%
\begin{eqnarray}
\Lambda _{\alpha \beta \mu }(t,\tau ) &=&+\mathrm{Tr}_{e}[\mathcal{F}%
_{\alpha }(\tau )\mathcal{F}_{\beta }(t)\rho _{0}^{e}]\ \mathrm{Tr}_{e}[%
\mathcal{F}_{\mu }(t)\rho _{0}^{e}] \\
&&-\mathrm{Tr}_{e}[\mathcal{F}_{\alpha }(\tau )\mathcal{F}_{\mu }(t)\rho
_{0}^{e}]\ \mathrm{Tr}_{e}[\mathcal{F}_{\beta }(t)\rho _{0}^{e}],  \notag
\end{eqnarray}%
which only depends on the initial environment state $\rho _{0}^{e}$ and
environment superoperators $\{\mathcal{F}_{\alpha }(t)\}.$ The probability $%
P(\breve{y})$ is%
\begin{equation}
P(\breve{y})=\sum_{\alpha }\langle \breve{y}|\mathbb{S}_{\alpha }[\rho _{%
\underline{\mathbf{x}}}]|\breve{y}\rangle \mathrm{Tr}_{e}[\mathcal{F}%
_{\alpha }(t)\rho _{0}^{e}].
\end{equation}%
The previous formula give an exact analytical expression for the CPF
correlation that is valid for a broad class of problems (see Sec. IV below).

\subsubsection{Random scheme}

In the random scheme, $P(z,\breve{y},x)$ reads~\cite{BIF}%
\begin{equation}
\frac{P(z,\breve{y},x)}{P(x)}\overset{r}{=}\mathrm{Tr}_{se}(E_{z}\mathcal{G}%
_{t+\tau ,t}^{se}[\rho _{\breve{y}}\otimes \mathrm{Tr}_{s}(\mathcal{G}%
_{t,0}^{se}[\rho _{x}^{se}])])\wp (\breve{y}|x),
\end{equation}%
where as before $E_{m}\equiv |m\rangle \langle m|,$ $\rho _{m}\equiv
|m\rangle \langle m|$ $[m=z,\breve{y},x],$ and $\rho _{x}^{se}\equiv \rho
_{x}\otimes \rho _{0}^{e},$ while $P(x)=\langle x|\rho _{0}^{s}|x\rangle .$
Furthermore, the conditional probability $\wp (\breve{y}|x)$ can be freely
chosen. Using the propagator expression~(\ref{SEPropagator}), it follows%
\begin{equation}
\frac{P(z,\check{y},x)}{P(x)}\overset{r}{=}\sum_{\alpha }\langle z|\mathbb{S}%
_{\alpha }[\rho _{\check{y}}]|z\rangle \mathrm{Tr}_{e}[\mathcal{F}_{\alpha
}(\tau )\rho _{t}^{e}]\wp (\check{y}|x).  \label{P3Random}
\end{equation}%
Consequently, independently of the chosen measurement observables, it is
confirmed that the CPF correlation [Eq.~(\ref{CPFexplicit})] vanishes
identically in this scheme,%
\begin{equation}
C_{pf}(t,\tau )|_{\breve{y}}\overset{r}{=}0.  \label{CPFRandomCero}
\end{equation}%
In fact, the sum term $[\sum_{\alpha }\cdots ]$ in Eq.~(\ref{P3Random}) can
be read as $P(z|\check{y}),$ leading to the Markovian structure $P(z,\check{y%
},x)=$ $P(z|\check{y})\wp (\check{y}|x)P(x)\rightarrow P(z,x|\check{y})=P(z|%
\check{y})P(x|\check{y}).$

\section{Examples}

In order to exemplify the developed results, we consider different single
and multipartite system dynamics interacting with a quantum casual bystander
environment.

\subsection{Single qubit system}

The system is a qubit, while the quantum degrees of freedom of the
environment correspond to a two-level fluorescent system with decay rate $%
\gamma $ and Rabi frequency $\Omega $ \cite{carmichaelbook}. Their mutual
evolution [Eq.~(\ref{GeneralSU})] is written as%
\begin{equation}
\frac{d}{dt}\rho _{t}^{se}=-i\frac{\Omega }{2}[\sigma _{x},\rho
_{t}^{se}]+\gamma \Big{(}\sigma \mathbb{S}[\rho _{t}^{se}]\sigma ^{\dag }-%
\frac{1}{2}\{\sigma ^{\dag }\sigma ,\rho _{t}^{se}\}_{+}\Big{)}.
\label{Fluor}
\end{equation}%
The operators $\sigma _{x},$ $\sigma ^{\dag },$ and $\sigma $ are
respectively the $x$-Pauli matrix and the raising and lowering operators in
the two-dimensional environment Hilbert space $\mathcal{H}_{e}.$ The unique
system contribution is the superoperator [see Eq.~(\ref{Kraus})]%
\begin{equation}
\mathbb{S}[\bullet ]=\sigma _{z}\mathbb{[}\bullet ]\sigma _{z},
\end{equation}%
where $\sigma _{z}$ is the $z$-Pauli matrix in the system Hilbert space $%
\mathcal{H}_{s}.$ Thus, the system is subjected to a dephasing process
driven by the transitions of the fluorescent system.

\subsubsection{System-environment propagator}

Considering (bipartite) uncorrelated initial conditions, $\rho
_{0}^{se}=\rho _{0}^{s}\otimes \rho _{0}^{e},$ the propagator of Eq.~(\ref%
{Fluor}) can be written with the structure~(\ref{SEPropagator}),%
\begin{equation}
\rho _{t}^{se}=\rho _{0}^{s}\otimes \mathcal{F}_{+}(t)[\rho _{0}^{e}]+%
\mathbb{S}[\rho _{0}^{s}]\otimes \mathcal{F}_{-}(t)[\rho _{0}^{e}].
\label{FluorRhoSEsolution}
\end{equation}%
At any time, this bipartite state is a separable one [Eq.~(\ref%
{FormalSolution})]. Consistently, a positive partial transpose criterion 
\cite{entanglement} is fulfilled. On the other hand, the environment
superoperators $\mathcal{F}_{\pm }$ can be written as%
\begin{equation}
\mathcal{F}_{\pm }(t)[\bullet ]=\frac{1}{2}(\mathcal{G}_{t,0}^{+}[\bullet
]\pm \mathcal{G}_{t,0}^{-}[\bullet ]),
\end{equation}%
where the auxiliary superoperators $\mathcal{G}_{t}^{\pm }[\bullet ]$ are
defined by the evolutions,%
\begin{equation}
\frac{d}{dt}\mathcal{G}_{t,t_{0}}^{\pm }\!\!=\!\!-i\frac{\Omega }{2}[\sigma
_{x},\mathcal{G}_{t,t_{0}}^{\pm }]+\gamma \Big{(}\pm \sigma \mathcal{G}%
_{t,t_{0}}^{\pm }\sigma ^{\dag }-\frac{1}{2}\{\sigma ^{\dag }\sigma ,%
\mathcal{G}_{t,t_{0}}^{\pm }\}_{+}\Big{)},  \label{GPM}
\end{equation}%
with the initial conditions $\mathcal{G}_{t_{0,}t_{0}}^{\pm }=\mathrm{I}%
_{e}. $ These equations can be solved in an analytical way.

Both superoperators $\mathcal{G}_{t,t_{0}}^{\pm }$ define the system and
environment dynamics,%
\begin{equation}
\rho _{t}^{e}=\mathcal{G}_{t,0}^{+}[\rho _{0}^{e}],\ \ \ \ \ \ \ \ f(\tau
|t)\equiv \mathrm{Tr}_{e}(\mathcal{G}_{t+\tau ,t}^{-}[\rho _{t}^{e}]).
\label{Coherence}
\end{equation}%
In fact, from Eqs.~(\ref{Fluor}) and (\ref{GPM}) it is simple to realize
that $\mathcal{G}_{t,t_{0}}^{+}$ is the propagator of the environment
degrees of freedom. Furthermore, from Eq.~(\ref{FluorRhoSEsolution}) it is
simple to show that the function$\ f(\tau |t)$ sets the system coherence
decay,%
\begin{equation}
\rho _{t}^{s}=\left( 
\begin{array}{cc}
p_{up} & f(t|0)c_{up} \\ 
f(t|0)c_{dn} & p_{dn}%
\end{array}%
\right) ,  \label{TLSDensity}
\end{equation}%
where $p_{up}$ and $p_{dn}$ are the initial upper and lower system
populations while the nondiagonal contributions $c_{up}$ and $c_{dn}$ are
the initial system coherences. Furthermore,%
\begin{equation}
f(\tau |t)=e^{-\gamma \tau }a_{t}+e^{-\gamma \tau /4}\Big{[}b_{t}\cosh
(\Delta \tau )+c_{t}\frac{\sinh (\Delta \tau )}{\Delta }\Big{]},
\label{Coherol}
\end{equation}%
where for shortening the expression we introduced the coefficient\ $\Delta
\equiv \sqrt{(\gamma /4)^{2}-\Omega ^{2}}.$ Explicit expressions for the
time-dependent coefficients $a_{t},$\ $b_{t},$ and $c_{t},$ can be found in
the Appendix.

\subsubsection{Operator correlations}

From Eq. (\ref{FluorRhoSEsolution}), using that $\lim_{t\rightarrow \infty }%
\mathcal{G}_{t,t_{0}}^{+}[\rho _{0}^{e}]=\rho _{\infty }^{e}$ and $%
\lim_{t\rightarrow \infty }\mathcal{G}_{t,t_{0}}^{-}[\rho _{0}^{e}]=0,$ it
follows%
\begin{equation}
\lim\nolimits_{t\rightarrow \infty }\rho _{t}^{se}=\rho _{\mathrm{\infty }%
}^{s}\otimes \rho _{\mathrm{\infty }}^{e},\ \ \ \ \ \ \rho _{\mathrm{\infty }%
}^{s}=\frac{1}{2}(\rho _{0}^{s}+\mathbb{S}[\rho _{0}^{s}]),
\end{equation}%
where $\rho _{\mathrm{\infty }}^{e}$ is the stationary state of a two-level
fluorescent system (see Appendix). Thus, when the environment at the initial
time begins in its stationary state, the conditions~(\ref{QRTConditions})
are fulfilled, indicating the validity of the QRT in the stationary regime.
This conclusion is corroborated by the following explicit calculation of
operator expectation values and correlations. Nevertheless, we also found
that for some operator correlations, the QRT is valid at all times.

Introducing the vector of system operators $\mathbf{\sigma }\equiv \{\sigma
_{x},\sigma _{y},\sigma _{z}\},$ their expectation values, from Eq.~(\ref%
{OperatorBipartite}), read%
\begin{equation}
\overline{\mathbf{\sigma }(\tau )}=\boldsymbol{\hat{T}}(\tau |0)\ \overline{%
\mathbf{\sigma }(0)},  \label{Operators}
\end{equation}%
where $\overline{\mathbf{\sigma }(0)}=\mathrm{Tr}_{s}[\mathbf{\sigma }%
(0)\rho _{0}^{s}].$ Similarly, for the Pauli operators there are nine
possible correlations $\overline{\sigma _{i}(t)\sigma _{j}(t+\tau )}.$ For
six of them, from Eq.~(\ref{Correlations}) we get 
\begin{subequations}
\label{CorreXYQRT}
\begin{eqnarray}
\overline{\sigma _{x}(t)\mathbf{\sigma }(t+\tau )} &=&\boldsymbol{\hat{T}}%
(\tau |t)\ \overline{\sigma _{x}(t)\mathbf{\sigma }(t)}, \\
\overline{\sigma _{y}(t)\mathbf{\sigma }(t+\tau )} &=&\boldsymbol{\hat{T}}%
(\tau |t)\ \overline{\sigma _{y}(t)\mathbf{\sigma }(t)}.
\end{eqnarray}%
In these expressions, the matrix $\boldsymbol{\hat{T}}(\tau |t)$ reads 
\end{subequations}
\begin{equation}
\boldsymbol{\hat{T}}(\tau |t)=\mathrm{diag}\{f(\tau |t),f(\tau |t),1\}.
\end{equation}%
When the environment begins in its stationary state $\rho _{0}^{e}=\rho
_{\infty }^{e}=\lim_{t\rightarrow }\rho _{t}^{e},$ it follows that $f(\tau
|t)=f(\tau |0)$ [see Eq.~(\ref{Coherence})], and consequently $\boldsymbol{%
\hat{T}}(\tau |t)=\boldsymbol{\hat{T}}(\tau |0).$ Thus, the six
correlations~(\ref{CorreXYQRT}) \textit{at any time} $(t$ and $\tau )$
evolve as the expectation values [Eq.~(\ref{Operators})] indicating the
absence of any departure with respect to the QRT.

The unique correlations [Eq.~(\ref{Correlations})] that depart \ (at any
finite time) from the predictions of the QRT are%
\begin{equation}
\overline{\sigma _{z}(t)\mathbf{\sigma }(t+\tau )}=\boldsymbol{\tilde{T}}%
(\tau ,t)\overline{\sigma _{z}(t)\mathbf{\sigma }(t)},  \label{Zetales}
\end{equation}%
where the matrix $\boldsymbol{\tilde{T}}(\tau ,t)$ is%
\begin{equation}
\boldsymbol{\tilde{T}}(\tau ,t)=\mathrm{diag}\left\{ \frac{f(t+\tau |0)}{%
f(t|0)},\frac{f(t+\tau |0)}{f(t|0)},1\right\} .
\end{equation}%
If the semigroup property $f(t+\tau |0)=f(\tau |t)f(t|0)$ becomes valid, the
QRT is recovered. This happens when the system coherence can be approximated
by an exponential decay behavior. On the other hand, using that $\overline{%
\sigma _{z}(t)\mathbf{\sigma }(t)}=\mathrm{diag}\{f(t|0),f(t|0),1\}\overline{%
\sigma _{z}(0)\mathbf{\sigma }(0)}$ we get $\lim_{t\rightarrow \infty }%
\overline{\sigma _{z}(t)\mathbf{\sigma }(t)}=\{0,0,1\}.$ Thus, consistently
with Eq.~(\ref{QRTConditions}), in the long time limit the correlations~(\ref%
{Zetales}) obey the same evolution as the operator expectation values [Eq.~(%
\ref{Operators})], $\lim_{t\rightarrow \infty }\overline{\sigma _{z}(t)%
\mathbf{\sigma }(t+\tau )}=\boldsymbol{\hat{T}}(\tau |0)\{0,0,1\},$
indicating the validity of the QRT in the stationary regime.

\subsubsection{Memory witnesses}

A deeper understanding of the memory effects developed in the studied model
can be achieved by comparing operational and non-operational memory
witnesses.

For \textit{non-operational approaches}, the central ingredient to analyze
is the system density matrix evolution. From Eq.~(\ref{TLSDensity}),
straightforwardly we get%
\begin{equation}
\frac{d}{dt}\rho _{t}^{s}=\frac{1}{2}\gamma _{t}(\mathbb{S}[\rho
_{t}^{s}]-\rho _{t}^{s}),\ \ \ \ \ \gamma _{t}=-\frac{d}{dt}\ln [f(t|0)],
\label{Rate}
\end{equation}%
where $f(t|0)$ follows from Eq.~(\ref{Coherol}). 
\begin{figure}[tbp]
\includegraphics[bb=25 594 760
1134,angle=0,width=8.5cm]{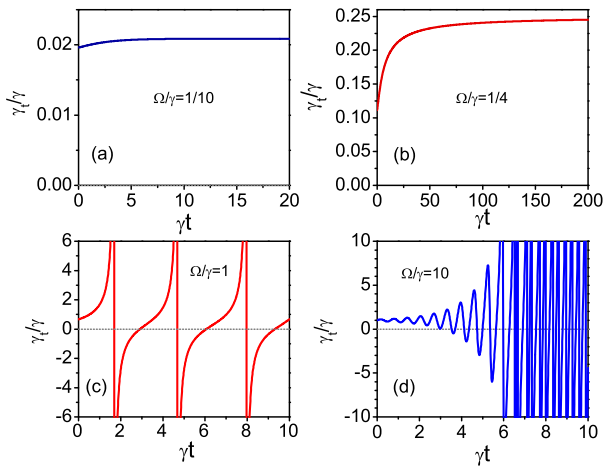}
\caption{Time dependence of the decoherence rate Eq.~(\protect\ref{Rate})
corresponding to the underlying dynamics (\protect\ref{Fluor}). The
environment begins in its stationary state. In each plot, the parameters are
(a) $\Omega /\protect\gamma =1/10,$ (b) $\Omega /\protect\gamma =1/4,$ (c) $%
\Omega /\protect\gamma =1,$ and (d) $\Omega /\protect\gamma =10.$ }
\end{figure}

The negativity of $\gamma _{t}$ can be used as an indicator of memory
effects~\cite{canonicalCresser}. We assume that the environment begins in
its stationary state, $\rho _{0}^{e}=\rho _{\infty }^{e}.$ Thus, $f(\tau
|t)=f(\tau |0)$ [see Eq.~(\ref{Coherence})]. In Fig.~1 we plot $\gamma _{t}$
for different values of the quotient $\Omega /\gamma .$ For $\Omega /\gamma
\leq 1/4$ the rate is always positive, while for $\Omega /\gamma >1/4$ it
develops periodical divergences. Thus, the dynamics is non-Markovian in this
last regime. The same conclusion follows from the trace distance between two
initial states~\cite{BreuerFirst}. On the other hand, for $\Omega /\gamma
\gg 1/4,$ the rate $\gamma _{t}$ approach a constant value, implying that a
Markovian regime is reached again.

The previous rate behaviors can be understood from the underlying
environment dynamics. For $\Omega /\gamma \ll 1/4,$ the probability
distribution of the elapsed time between environment (fluorescent)
transitions approach an exponential function with average time $[2\Omega
^{2}/\gamma ]^{-1}$ \cite{carmichaelbook}. Consequently, the coherence decay
function (induced by the application of the superoperator $\mathbb{S}$) can
be approximated as $f(\tau |0)\approx \exp [-t(2\Omega ^{2}/\gamma )],$
which implies $\gamma _{t}\approx 2\Omega ^{2}/\gamma $ [Fig.~1(a)]. This
regime changes drastically when $\Omega /\gamma =1/4$ [Fig.~1(b)], where the
environment starts to develop Rabi oscillations. Around $\Omega /\gamma
\approx 1,$ the system coherence $f(t|0)$ vanishes in a oscillatory way.
Consequently, $\gamma _{t}$ [Eq.~(\ref{Rate})] develops periodic divergences
[Fig.~1(c)]. For $\Omega /\gamma \gg 1,$ the effect of the (fast)
environment Rabi oscillations over the system cancel out in average, leading
to the coherence decay $f(\tau |0)\approx \exp (-t\gamma ).$ Thus, $\gamma
_{t}\approx \gamma .$ This tendency is clearly seen in Fig.~1(d) at the
initial stage.

The previous dynamical regimes can be analyzed from the \textit{operational
approach}. For the dynamics~(\ref{Fluor}), using the explicit propagator
Eq.~(\ref{FluorRhoSEsolution}), all statistical objects that define the CPF
correlation can explicitly be evaluated. We consider that the three
consecutive measurements are performed in the $\hat{x}$-direction of the
system Bloch sphere. Thus, in all cases, the possible measurement outcomes
(eigenvalues) are $m=\pm 1$ $(m=x,y,z),$ while the corresponding
eigenvectors are $|m\rangle =|up\rangle +m|dn\rangle ,$ where $|up\rangle $
and $|dn\rangle $ are respectively the upper and down states of the system
[see Eq.~(\ref{TLSDensity})].

In the deterministic scheme, the joint outcome probability Eq.~(\ref{P3Deter}%
) becomes%
\begin{equation}
P(z,\breve{y},x)\overset{d}{=}\frac{1}{4}[1+z\breve{y}f(\tau |t)+zxf(t+\tau
|0)+\breve{y}xf(t|0)]P(x).
\end{equation}%
The CPF correlation Eq.~(\ref{CPFGen}) reads%
\begin{equation}
C_{pf}(t,\tau )|_{\breve{y}}\overset{d}{=}\frac{1-\langle x\rangle ^{2}}{4[P(%
\breve{y})]^{2}}[f(t+\tau |0)-f(\tau |t)\ f(t|0)],  \label{CPFDet}
\end{equation}%
with $P(\breve{y})=[1+\breve{y}\langle x\rangle f(t|0)]/2,$ jointly with $%
\langle x\rangle =\sum_{x=\pm 1}xP(x)$ and $P(x)=\langle x|\rho
_{0}^{s}|x\rangle $ where $\{|x\rangle \}$ are the eigenvectors of the $\hat{%
x}$-Pauli matrix. 
\begin{figure}[tbp]
\includegraphics[bb=55 595 745
1155,angle=0,width=8.5cm]{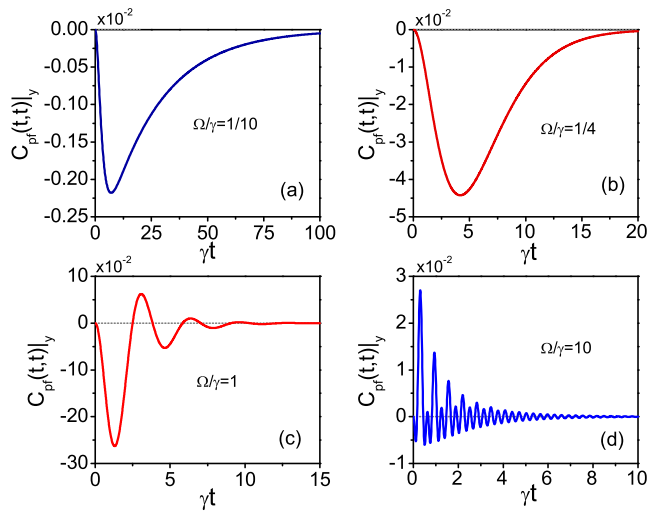}
\caption{CPF correlation [Eq.~(\protect\ref{CPFDet})] with equal time
intervals\ $\protect\tau =t$ corresponding to the system-environment\ model (%
\protect\ref{Fluor}). The parameters $\Omega /\protect\gamma $ are the same
than in Fig. 1. Similarly, the environment begins in its stationary state.
In all cases, the system initial condition is such that $\langle x\rangle
=0. $}
\end{figure}

In Fig.~2, for the same parameter regimes shown in Fig.~1, we plot the CPF
correlation at equal time intervals, $\tau =t.$ The environment also begins
in its stationary state, $\rho _{0}^{e}=\rho _{\infty }^{e}.$

Contrarily to the non-operational memory witnesses, the CPF correlation
indicates\textit{\ the presence of memory effects for all parameter regimes}%
, even when the rate $\gamma _{t}$ is positive at all times. Consistently,
for $\Omega /\gamma \ll 1/4,$ the maximal absolute value of the CPF
correlation diminishes [Fig.~2(a)], indicating the proximity of a Markovian
regime. When $\Omega /\gamma =1/4$ [Fig.~2(b)], the CPF is negative at all
times and does not develop oscillations. For $\Omega /\gamma \approx 1,$ it
develops oscillations and its absolute value is maximal [Fig.~2(c)],
indicating strong memory effects. Consistently, when $\Omega /\gamma \gg 1,$
the CPF correlation oscillates but with a smaller amplitude [Fig.~2(d)],
indicating again the approaching of a Markov regime.

Contrarily to memory witnesses based only on the unperturbed system
dynamics, the CPF correlation indicates a Markovian regime only in the
limits $\Omega /\gamma \rightarrow 0$ and $\Omega /\gamma \rightarrow \infty
.$ In addition, the operational approach gives a much deeper
characterization when considering the random scheme. From Eq.~(\ref{P3Random}%
), for the joint probabilities we get%
\begin{equation}
P(z,\breve{y},x)\overset{r}{=}\frac{1}{2}[1+z\breve{y}f(\tau |t)]\wp (\breve{%
y}|x)P(x).
\end{equation}%
As expected, a Markovian property is fulfilled, leading consistently to $%
C_{pf}(t,\tau )|_{\breve{y}}\overset{r}{=}0$ [Eq.~(\ref{CPFRandomCero})] for
arbitrary initial environment states. This result indicates the presence of
a casual bystander environment, \textit{property that cannot be resolved
with non-operational approaches.}

\subsection{Multipartite qubit systems}

The developed formalism also allows to study the coupling of multipartite
systems with a casual bystander environment. In contrast to Eq.~(\ref{Fluor}%
), here we consider a set of $N$ qubits. For simplicity, we assume the
system-environment evolution%
\begin{eqnarray}
\frac{d}{dt}\rho _{t}^{se}\! &=&\!-i\frac{\Omega }{2}[\sigma _{x},\rho
_{t}^{se}]+\gamma \Big{(}\sigma \mathbb{S}_{\underline{\mathbf{a}}}[\rho
_{t}^{se}]\sigma ^{\dag }-\frac{1}{2}\{\sigma ^{\dag }\sigma ,\rho
_{t}^{se}\}_{+}\Big{)}  \notag \\
&&+\varphi \Big{(}\sigma ^{\dag }\mathbb{S}_{\underline{\mathbf{b}}}[\rho
_{t}^{se}]\sigma -\frac{1}{2}\{\sigma \sigma ^{\dag },\rho _{t}^{se}\}_{+}%
\Big{)},  \label{Multipartito}
\end{eqnarray}%
As before, $\sigma _{x},$ $\sigma ^{\dag },$ and $\sigma $ are respectively
the $x$-Pauli matrix and the raising and lowering operators in the
two-dimensional environment Hilbert space $\mathcal{H}_{e}.$ Thus, the
environment corresponds to a two-level fluorescent-like system with Rabi
frequency $\Omega ,$\ decay rate $\gamma ,$ while the rate $\varphi $\
scales the presence of thermally induced excitations \cite{breuerbook}.

Each of the system superoperators $\mathbb{S}_{\mathbf{\alpha }}[\bullet
]\equiv \sigma _{\mathbf{\alpha }}\bullet \sigma _{\mathbf{\alpha }}$ $(%
\mathbf{\alpha }=\underline{\mathbf{a}},\underline{\mathbf{b}})$ are defined
by an arbitrary (multipartite) Pauli string $\sigma _{\mathbf{\alpha }%
}=\sigma _{\alpha _{1}}\otimes \cdots \otimes \sigma _{\alpha _{N}},$ which
consists in the external product of $N$ arbitrary Pauli operators acting on
each qubit. These superoperators are applied over the system whenever the
environment suffers a transition between its (two) states. From Eq.~(\ref%
{Multipartito}), it follows that $\mathbb{S}_{\underline{\mathbf{a}}}$ is
applied when an environmental transition between the upper and lower states
occurs, while $\mathbb{S}_{\underline{\mathbf{b}}}$ is applied for the
inverse (thermally-induced) transition.

The bipartite propagator associated to Eq.~(\ref{Multipartito}) can be
written with the structure Eq.~(\ref{SEPropagator}). The label of the
superoperators $\{\mathcal{F}_{\mathbf{\alpha }}(t)\}$ runs over the values $%
\mathbf{\alpha }=\mathbf{0},\underline{\mathbf{a}},\underline{\mathbf{b}},%
\underline{\mathbf{c}},$ where $\mathbb{S}_{\underline{\mathbf{c}}}=\mathbb{S%
}_{\underline{\mathbf{b}}}\mathbb{S}_{\underline{\mathbf{a}}}.$ The
calculations that lead to explicit solutions for the environment
superoperators $\{\mathcal{F}_{\mathbf{\alpha }}(t)\}$ are presented in the
Appendix. From these expressions and Eq.~(\ref{SEPropagator}), it follows
that%
\begin{equation}
\lim\nolimits_{t\rightarrow \infty }\rho _{t}^{se}=\rho _{\mathrm{\infty }%
}^{s}\otimes \rho _{\mathrm{\infty }}^{e},\ \ \ \ \ \rho _{\mathrm{\infty }%
}^{s}=\frac{1}{4}(\rho _{0}^{s}+\sum_{\alpha }\mathbb{S}_{\alpha }[\rho
_{0}^{s}]),
\end{equation}%
where $\rho _{\mathrm{\infty }}^{s}$ is the multipartite system stationary
state while the (two-level) state $\rho _{\mathrm{\infty }}^{e}$ follows by
tracing out the system degrees of freedom in Eq.~(\ref{Multipartito}). In
consequence, as in the previous example, the QRT is valid for stationary
correlations of the system.

From Eq.~(\ref{SEPropagator}) the system state at any time, $\rho _{t}^{s}=%
\mathrm{Tr}_{e}(\rho _{t}^{se}),$ can straightforwardly be written as a
statistical superposition of Kraus maps,%
\begin{equation}
\rho _{t}^{s}=p_{t}^{\mathbf{0}}\rho _{0}+\sum_{\mathbf{\alpha }=\underline{%
\mathbf{a}},\underline{\mathbf{b}},\underline{\mathbf{c}},}p_{t}^{\mathbf{%
\alpha }}\mathbb{S}_{\mathbf{\alpha }}[\rho _{0}],
\end{equation}%
where the weights are $p_{t}^{\mathbf{\alpha }}\equiv \mathrm{Tr}_{e}[%
\mathcal{F}_{\alpha }(t)[\rho _{0}^{e}]].$ Similarly, the density matrix
evolution can be written as%
\begin{equation}
\frac{d\rho _{t}^{s}}{dt}=\sum_{\mathbf{\alpha }=\underline{\mathbf{a}},%
\underline{\mathbf{b}},\underline{\mathbf{c}},}\gamma _{t}^{\mathbf{\alpha }%
}(\mathbb{S}_{\mathbf{\alpha }}[\rho _{t}^{s}]-\rho _{t}^{s}).
\label{MasterN}
\end{equation}%
Simple expressions for the probabilities $\{p_{t}^{\mathbf{\alpha }}\}$ and
rates $\{\gamma _{t}^{\mathbf{\alpha }}\}$ are obtained when $\varphi
=\gamma $ in Eq.~(\ref{Multipartito}). We get 
\begin{subequations}
\begin{eqnarray}
p_{t}^{\mathbf{0}} &=&\frac{1}{2}e^{-\gamma t}\Big{[}\cosh (\gamma t)+\frac{%
\gamma ^{2}}{\chi ^{2}}\cos (\chi t)+\frac{\Omega ^{2}}{\chi ^{2}}\Big{]}, \\
p_{t}^{\underline{\mathbf{a}}} &=&p_{t}^{\underline{\mathbf{b}}}=\frac{1}{4}%
[1-e^{-2\gamma t}], \\
p_{t}^{\underline{\mathbf{c}}} &=&\frac{1}{2}e^{-\gamma t}\Big{[}\cosh
(\gamma t)-\frac{\gamma ^{2}}{\chi ^{2}}\cos (\chi t)-\frac{\Omega ^{2}}{%
\chi ^{2}}\Big{]},
\end{eqnarray}%
where $\chi \equiv \sqrt{\gamma ^{2}+\Omega ^{2}}.$ From these expressions,
the rates in Eq.~(\ref{MasterN}) are 
\end{subequations}
\begin{equation}
\gamma _{t}^{\underline{\mathbf{a}}}=\gamma _{t}^{\underline{\mathbf{b}}}=%
\frac{1}{2}\gamma ,\ \ \ \ \ \ \ \gamma _{t}^{\underline{\mathbf{c}}}=\frac{1%
}{2}\frac{\chi \sin (\chi t)}{(\Omega /\gamma )^{2}+\cos (\chi t)}.
\label{TanTrig}
\end{equation}%
Notice that $\gamma _{t}^{\underline{\mathbf{c}}}$ presents an oscillatory
behavior at any time, which develops divergences only when $(\Omega /\gamma
)^{2}<1.$ When $\Omega =0,$ it reduces to $\gamma _{t}^{\underline{\mathbf{c}%
}}=(1/2)\gamma \tan (\gamma t),$ recovering the rates of the
\textquotedblleft trigonometric eternal non-Markovian\textquotedblright\
dynamics introduced in Ref.~\cite{arxiv}, where the environment dynamics is
an incoherent one. Thus, we can read Eq.~(\ref{Multipartito}) as a \textit{%
quantum (coherent) generalization} $(\Omega \neq 0)$ of the incoherent
environment studied in \cite{arxiv}.

The CPF correlation can also be obtained in the present case. Assuming that
the three measurements correspond to the observable $\sigma _{\mathbf{\alpha 
}}$ $(\alpha =\underline{\mathbf{a}}$ or $\underline{\mathbf{b}}),$ from
Eq.~(\ref{CPFGen}) we get (with $\varphi =\gamma )$%
\begin{eqnarray}
&&\!\!\!\!\!\!\!\!C_{pf}(t,\tau )|_{\breve{y}}\overset{d}{=}-\frac{%
(1-\langle x\rangle ^{2})}{[2^{N}P(\breve{y})]^{2}}e^{-(t+\tau )\gamma /2}%
\Big{[}\frac{\gamma ^{2}}{\chi ^{2}}\sin (t\chi )\sin (\tau \chi )  \notag \\
&&\ \ \ \ \ \ \ \ \ \ \ \ \ \ -\frac{4\gamma ^{2}\Omega ^{2}}{\chi ^{4}}\sin
^{2}(t\chi /2)\sin ^{2}(\tau \chi /2)\Big{]},
\end{eqnarray}%
where as before $\chi =\sqrt{\gamma ^{2}+\Omega ^{2}},$ and $P(\breve{y}%
)=2^{-N}\{1+\breve{y}\langle x\rangle e^{-\gamma t}[\Omega ^{2}+\gamma
^{2}\cos (t\chi )]/\chi ^{2}\}.$ Furthermore, $\langle x\rangle
=\sum_{x}x\langle x|\rho _{0}^{s}|x\rangle ,$ where $\{x\}$ and $\{|x\rangle
\}$ are respectively the eigenvalues and eigenvectors associated to the
first measurement observable. When the three measurements correspond to the
observable $\sigma _{\underline{\mathbf{c}}},$ we get the accidental
vanishing $C_{pf}(t,\tau )|_{y}\overset{d}{=}0.$ On the other hand, in the
random scheme Eq.~(\ref{CPFRandomCero}) guarantee that $C_{pf}(t,\tau )|_{y}%
\overset{r}{=}0$ for any system observables and initial environment states.

\section{Summary and conclusions}

Quantum memory effects can be induced by environments whose state and
dynamical behavior are not affected at all by their interaction with the
system of interest. Based on a bipartite representation of the
system-environment dynamics, in this paper we have explored the most general
interaction structures that are consistent with this class of non-Markovian
casual bystander environments.

While unitary interactions must be discarded, we have found the most general
dissipative coupling structures [Eq.~(\ref{GeneralSU})] that are consistent
with the demanded constraint. The degrees of freedom associated to the
environment are governed by a Lindblad evolution. The corresponding system
dynamic turns out to be defined by a set of arbitrary completely positive
transformations whose action is conditioned to the environment dynamics.

The bipartite system-environment state can always be written as a separable
one [Eq.~(\ref{FormalSolution})], indicating the absence of quantum
entanglement between both parts. Nevertheless, in contrast to a purely
incoherent case, the environment may develop quantum coherent behaviors.
Consistently, by subjecting the degrees of freedom of the environment to a
continuous-in-time measurement process,\ a product state characterizes the
bipartite stochastic dynamics [Eq.~(\ref{StochasticRep})], where a
collisional dynamics defines the stochastic system evolution.

Similarly to incoherent environments, here the QRT is not valid in general.
Nevertheless, \textit{stationary} (system) operators correlations evolves in
the same way as expectation values when the bipartite system-environment
stationary state is an uncorrelated one [Eq.~(\ref{QRTConditions})].
Consequently, outside the stationary regime operator correlations can be
used as a witness of memory effects. Nevertheless, given that the absence of
stationary system-environment correlations may emerges in different models,
a deeper characterization of non-Markovianity can be achieved through an
operational approach.

The CPF correlation is an operational memory witness that relies on
performing three consecutive measurement processes over the system of
interest. This object was explicitly calculated in terms of a bipartite
propagator [Eq.~(\ref{SEPropagator})] associated to the studied
system-environment coupling. In a deterministic scheme, where the system
state is not modified after the intermediate measurement, the CPF
correlation detects departure with respect to a (probabilistic) Markovian
regime [Eq.~(\ref{CPFGen})]. In a random scheme, where the intermediate
post-measurement state is selected in a random way, the CPF correlation
vanishes when the environment is a casual bystander one [Eq.~(\ref%
{CPFRandomCero})]. This feature provides an explicit experimental procedure
for detecting when the studied properties apply.

All previous conclusions were supported by the explicit study of single and
multipartite qubits dynamics. The developed approach furnishes a solid basis
for constructing alternative underlying mechanisms that lead to quantum
memory effects. On the other hand, added to incoherent environments with a
classical self fluctuating dynamics, the studied dynamics define the most
general situation where quantum memory effects are not endowed with a
physical environment-to-system backflow of information. While unitary
system-environment interactions were discarded, the present results motivate
us to ask about different dynamical regimes where an effective non-Markovian
casual bystander environmental action could be recovered.

\section*{Acknowledgments}

The author thanks to Mariano Bonifacio for a critical reading of the
manuscript. This paper was supported by Consejo Nacional de Investigaciones
Cient\'{\i}ficas y T\'{e}cnicas (CONICET), Argentina.

\appendix

\section{Auxiliary expressions and calculus details}

Auxiliary expressions and calculation details are provided.

\subsection{Coefficients of the coherence decay function}

The coherence decay function $f(\tau |t)\equiv \mathrm{Tr}_{e}(\mathcal{G}%
_{t+\tau ,t}^{-}[\rho _{t}^{e}]),$ where $\mathcal{G}_{t+\tau ,t}^{-}$ is
defined by the evolution~(\ref{GPM}), can be written as in Eq.~(\ref{Coherol}%
), where the time-dependent coefficients are 
\begin{subequations}
\begin{eqnarray}
a_{t} &=&\left[ 1-\frac{2\gamma \Omega \overline{\sigma _{y}(t)}-\overline{%
\sigma _{z}(t)}\gamma ^{2}}{\gamma ^{2}+2\Omega ^{2}}\right] , \\
b_{t} &=&\frac{2\gamma \Omega \overline{\sigma _{y}(t)}-\overline{\sigma
_{z}(t)}\gamma ^{2}}{\gamma ^{2}+2\Omega ^{2}}, \\
c_{t} &=&-\frac{6\gamma \Omega \overline{\sigma _{y}(t)}+\overline{\sigma
_{z}(t)}(\gamma ^{2}+8\Omega ^{2})}{4(\gamma ^{2}+2\Omega ^{2})}.
\end{eqnarray}%
The overbar symbol denotes the expectation values $\overline{\sigma _{i}(t)}=%
\mathrm{Tr}_{e}[\rho _{t}^{e}\sigma _{i}],$ where $\sigma _{i}$ are Pauli
operators in $\mathcal{H}_{e}.$ The environment state follows from $\rho
_{t}^{e}=\mathcal{G}_{t,0}^{+}[\rho _{0}^{e}],$ where $\mathcal{G}_{t,0}^{+}$
is also defined by the evolution~(\ref{GPM}). The stationary environment
state $\rho _{\mathrm{\infty }}^{e}=\lim_{t\rightarrow \infty }\rho _{t}^{e}$
reads 
\end{subequations}
\begin{equation}
\rho _{\mathrm{\infty }}^{e}=\frac{1}{\gamma ^{2}+2\Omega ^{2}}\left( 
\begin{array}{cc}
\Omega ^{2} & -i\gamma \Omega \\ 
+i\gamma \Omega & \gamma ^{2}+\Omega ^{2}%
\end{array}%
\right) .
\end{equation}%
Assuming that the environment begins in this state, the previous
expectations values $\overline{\sigma _{i}(\infty )}=\lim_{t\rightarrow
\infty }\overline{\sigma _{i}(t)}=\mathrm{Tr}_{e}[\rho _{\infty }^{e}\sigma
_{i}]$ follows straightforwardly,%
\begin{equation}
\overline{\sigma _{y}(\infty )}=\frac{2\gamma \Omega }{\gamma ^{2}+2\Omega
^{2}},\ \ \ \ \ \ \ \overline{\sigma _{z}(\infty )}=-\frac{\gamma ^{2}}{%
\gamma ^{2}+2\Omega ^{2}}.
\end{equation}

\subsection{Multipartite system-environment propagator}

The bipartite propagator corresponding to the model (\ref{Multipartito}) can
be written as in Eq. (\ref{SEPropagator}). The solution for the set of
environment superoperators $\{\mathcal{F}_{\mathbf{\alpha }}(t)\}$ can be
obtained by defining the vector%
\begin{equation}
\mathcal{F}\equiv \{\mathcal{F}_{\mathbf{0}}(t),\mathcal{F}_{\underline{%
\mathbf{a}}}(t),\mathcal{F}_{\underline{\mathbf{b}}}(t),\mathcal{F}_{%
\underline{\mathbf{c}}}(t)\}.
\end{equation}%
It is written as $\mathcal{F}=(1/4)H\cdot \mathcal{G},$ where $H$ is a
four-dimensional Hadamard matrix, $H=\{\{1,1,1,1\},\{1,1,-1,-1\},\{1,-1,1,-1%
\},\{1,-1-,1,1\}\}.$ The components of the vector $\mathcal{G}$ are denoted
as%
\begin{equation}
\mathcal{G}\equiv \{\mathcal{G}_{t,0}^{++},\mathcal{G}_{t,0}^{+-},\mathcal{G}%
_{t,0}^{-+},\mathcal{G}_{t,0}^{--}\},
\end{equation}%
which in turn can be written as $\mathcal{G}=H\cdot \mathcal{F}.$ With these
definitions, the underlying model~(\ref{Multipartito}) implies the
time-evolutions%
\begin{eqnarray}
\frac{d\mathcal{G}^{uv}}{dt}\!\! &=&\!\!-i\frac{\Omega }{2}[\sigma _{x},%
\mathcal{G}^{uv}]\!-\!\frac{\gamma }{2}\{\sigma ^{\dag }\sigma ,\mathcal{G}%
^{uv}\}_{+}\!-\!\frac{\varphi }{2}\{\sigma \sigma ^{\dag },\mathcal{G}%
^{uv}\}_{+}  \notag \\
&&\!+u(\gamma \sigma \mathcal{G}^{uv}\sigma ^{\dag })+v(\varphi \sigma
^{\dag }\mathcal{G}^{uv}\sigma ),
\end{eqnarray}%
with initial conditions $\mathcal{G}_{t_{0},t_{0}}^{uv}=\mathrm{I}_{e}.$ For
shortening the expressions, we denoted $\mathcal{G}_{t,t_{0}}^{uv}%
\leftrightarrow \mathcal{G}^{uv}.$ The supra indexes are $u=\pm $\ and $%
v=\pm .$ The explicit analytical expressions for the four superoperators $%
\mathcal{G}^{uv}$ can be obtained by solving their evolution via Laplace
transform techniques.

\end{document}